\documentclass[a4paper,10pt,twoside]{cpc-hepnp}

\usepackage{multicol}
\usepackage{graphicx}
\usepackage{booktabs}
\usepackage{amssymb,bm,mathrsfs,bbm,amscd}
\usepackage[tbtags]{amsmath}
\usepackage{lastpage}

\begin{document}

\fancyhead[co]{\footnotesize N. Su et al: Hard-thermal-loop QED thermodynamics}

\footnotetext[0]{Received XX November 2009}

\title{Hard-thermal-loop QED thermodynamics}

\author{%
      Nan Su$^{1;1)}$\email{nansu@fias.uni-frankfurt.de}%
\quad Jens O. Andersen$^{2;2)}$\email{andersen@tf.phys.ntnu.no}%
\quad Michael Strickland$^{3,1;3)}$\email{mstrickl@gettysburg.edu}
}
\maketitle

\address{%
1~(Frankfurt Institute for Advanced Studies, D-60438 Frankfurt am Main, Germany)\\
2~(Department of Physics, Norwegian University of Science and Technology, N-7491 Trondheim, Norway)\\
3~(Department of Physics, Gettysburg College, Gettysburg, PA 17325, USA)\\
}

\begin{abstract}

The weak-coupling expansion for thermodynamic quantities in thermal field theories is poorly convergent unless the coupling constant is tiny. We discuss the calculation of the free energy for a hot gas of electrons and photons to three-loop order using hard-thermal-loop perturbation theory (HTLpt). We show that the hard-thermal-loop perturbation reorganization improves the convergence of the successive approximations to the QED free energy at large coupling, $e \sim 2$. The reorganization is gauge invariant by construction, and due to the cancellations among various contributions, we obtain a completely analytic result for the resummed thermodynamic potential at three loops.

\end{abstract}



\begin{multicols}{2}

\section{Introduction}

The calculation of thermodynamic functions for finite temperature field theories has a long history. In the early 1990s the free energy was calculated to order $g^4$ for massless scalar $\phi^4$ theory\cite{Frenkel:1992az,AZ-95}, QED\cite{Parwani:1994xi,AZ-95} and QCD\cite{AZ-95} respectively. The corresponding calculations to order $g^5$ were soon obtained afterwards\cite{Parwani:1994zz, Braaten:1995cm}\cite{Parwani:1994je,Andersen:1995ej}\cite{Zhai:1995ac,BN-96}. Recent results have extended the calculation of the QCD free energy by determining the coefficient of the $g\log g$ contribution\cite{Kajantie:2002wa}. For massless scalar theories the perturbative free energy is now known to order $g^6$\cite{Gynther:2007bw} and $g^8 \log g$\cite{Andersen:2009ct}.

Unfortunately, for all the above-mentioned theories the resulting weak-coupling approximations, truncated order-by-order in the coupling constant, are poorly convergent unless the coupling constant is tiny. In this proceedings we shall focus on the discussion of QED. Fig.~\ref{fig:pertpressure} shows the successive perturbative approximations to the QED free energy. As can be seen from this figure, at couplings larger than $e \sim 1$ the QED weak-coupling approximations exhibit poor convergence. To improve the bad convergence of perturbative expansions, several systematic resummation techniques have been introduced and they are summarized in references\cite{Blaizot:2003tw,Kraemmer:2003gd,Andersen:2004fp}. In the following we will discuss recent advances in the application of hard-thermal-loop perturbation theory (HTLpt) \cite{htl1,AS-01,htl2}.

\begin{center}
\includegraphics[width=7.5cm]{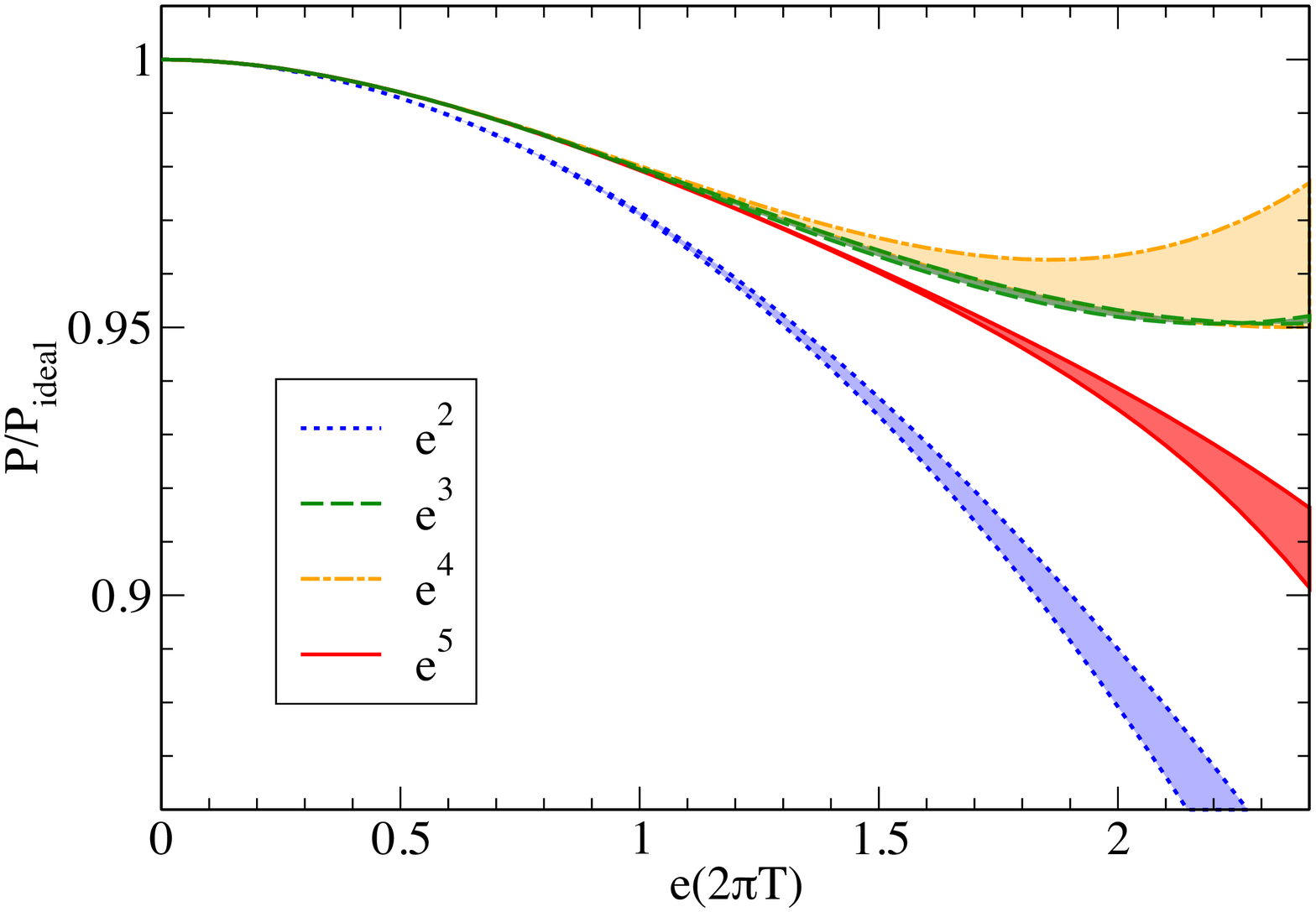}
\vspace{-2mm}
\figcaption{\label{fig:pertpressure} Successive perturbative approximations to the QED pressure (negative of the free energy). Each band corresponds to a truncated weak-coupling expansion to order $e^2$, $e^3$, $e^4$, and $e^5$, respectively. Shaded bands correspond to variation of the renormalization scale $\mu$ between $\pi T$ and $4 \pi T$.}
\end{center}

\end{multicols}
\begin{center}
\includegraphics[width=7.5cm]{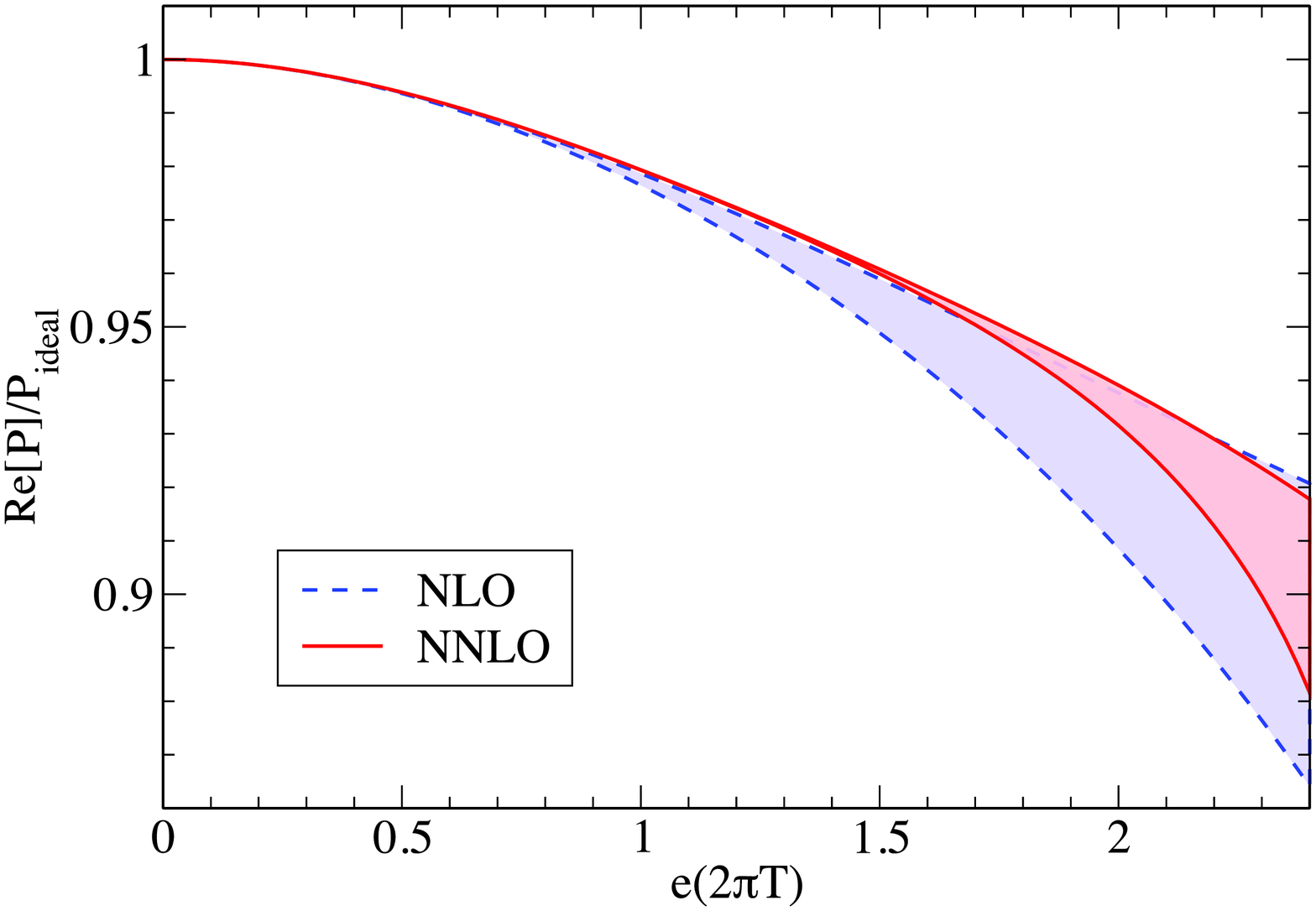}$\;\;\;\;$\includegraphics[width=7.5cm]{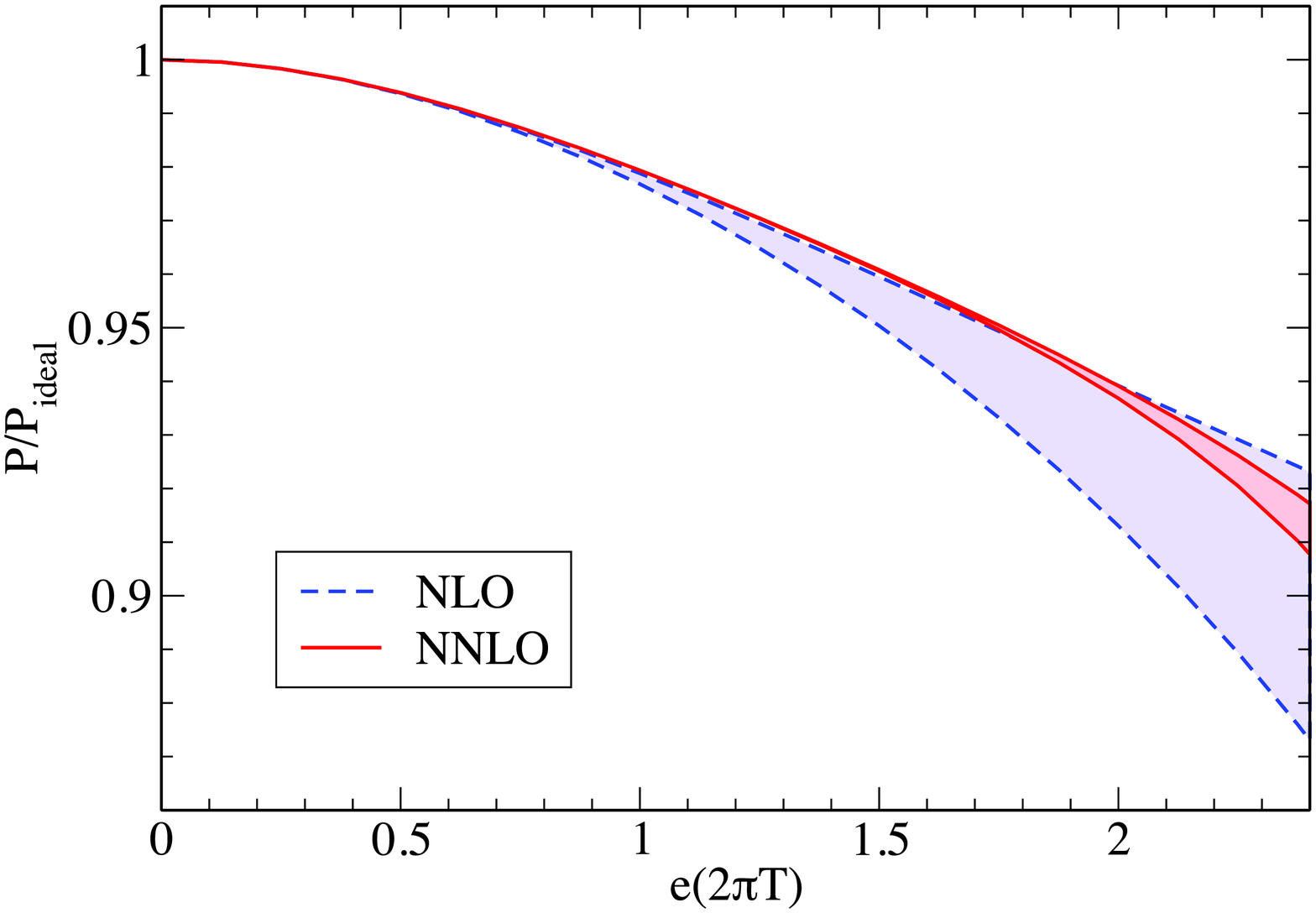}
\figcaption{\label{fig:NLONNLO} A comparison of the renormalization scale variations between NLO and NNLO HTLpt predictions for the free energy of QED with $N_f=1$ and the variational thermal masses (left) and the perturbative thermal masses (right). The bands correspond to varying the renormalization scale $\mu$ by a factor of 2 around $\mu=2\pi T$.}
\end{center}
\begin{multicols}{2}

\section{Three-loop hard-thermal-loop perturbation theory}

Hard-thermal-loop perturbation theory is inspired by variational perturbation theory\cite{yuk,steve,kleinert,deltaexp} and is a gauge-invariant extension of screened perturbation theory\cite{K-P-P-97,CK-98,spt,Andersen:2008bz}. The basic idea of the technique is to add and subtract an effective mass term from the bare Lagrangian and to associate the added piece with the free Lagrangian and the subtracted piece with the interactions. However, in gauge theories, one cannot simply add and subtract a local mass term since this would violate gauge invariance. Instead one adds and subtracts an HTL improvement term which modifies the propagators and vertices in such a way that the framework is manifestly gauge-invariant.

HTLpt has recently been pushed to three loops or the next-to-next-to-leading order (NNLO) and the details of the formalism and calculations are presented in Ref.\cite{Andersen:2009tw}. Here only a few selected results are listed.

With rescaled dimensionless parameters $\hat m_{D} = m_{D} /( 2 \pi T)$, $\hat m_{f} = m_{f} /( 2 \pi T)$, and $\hat\mu = \mu /( 2 \pi T)$, the renormalized NNLO thermodynamic potential reads
\end{multicols}
\ruleup
\begin{eqnarray}
\Omega_{\rm NNLO}&=&
- {\pi^2 T^4\over45} \Bigg\{
	1 + {7\over4}N_f - {15\over4} \hat m_D^3
	+ N_f {\alpha\over\pi} \Bigg[ -{25\over8}
	+ {15\over2} \hat m_D
	+15 \left(\log{\hat\mu \over 2}-{1\over2}
+\gamma+2\log2\right)\!\!\hat m_D^3
	- 90\hat m_D \hat m_f^2 \Bigg]
\nonumber \\ &&\hspace{0mm}
+ N_f \left({\alpha\over\pi}\right)^2
\Bigg[{15\over64}(35-32\log2)-{45\over2} \hat m_D\Bigg]
+ N_f^2 \left({\alpha\over\pi}\right)^2 \Bigg[{5\over4}{1\over\hat m_D}+{30}{\hat{m}_f^2\over\hat{m}_D}
\nonumber \\ &&\hspace{0mm}
+{25\over12}\left(\log{\hat\mu \over 2}+{1\over20}+{3\over5}\gamma-{66\over25}\log2
+{4\over5}{\zeta^{\prime}(-1)\over\zeta(-1)}
-{2\over5}{\zeta^{\prime}(-3)\over\zeta(-3)}
\right)  - 15\left(\log{\hat\mu \over 2}-{1\over2}
+\gamma+2\log2\right)\!\!\hat m_D
\Bigg]
\Bigg\} \;.
\label{Omega-NNLO}
\end{eqnarray}
\vspace{8mm}
\ruledown
\begin{multicols}{2}
There is also a corresponding next-to-leading order (NLO) thermodynamic potential that contains some numerical coefficients\cite{Andersen:2009tw}. We note that at NNLO all numerically determined coefficients drop out and we are left with a final result which is completely analytic.
 
The mass parameters $m_D$ and $m_f$ in HTLpt are in principle completely arbitrary. To complete a calculation, it is necessary to specify $m_D$ and $m_f$ as functions of $e$ and $T$. In Ref.\cite{Andersen:2009tw} we considered two possible mass prescriptions: 1) the variational thermal masses obtained from the gap equations; 2) the $e^5$ perturbative Debye mass\cite{Blaizot:1995kg,Andersen:1995ej} and the $e^3$ perturbative fermion mass\cite{carrington}. The resulting predictions for the free energy are shown in Fig.~\ref{fig:NLONNLO}. As can be seen from these figures both the variational and perturbative mass prescriptions seem to be consistent when going from NLO to NNLO. At the central value $\mu=2\pi T$, both prescriptions are the same to an accuracy of 0.6\% at $e=2.4$. As a further check, we show a comparison of our NNLO HTLpt results with a three-loop calculation obtained previously using a truncated three-loop $\Phi$-derivable approximation\cite{phijm} in Fig.~\ref{fig:PhivsNNLO}. As can be seen from this figure, there is very good agreement between the NNLO $\Phi$-derivable and HTLpt approaches even at large coupling.

\section{Conclusions and outlook}

In this proceedings we briefly discussed HTLpt, which is a gauge-invariant reorganization of finite temperature perturbation theory. We presented results of a recent three-loop HTLpt calculation of the QED free energy\cite{Andersen:2009tw} and showed that the HTLpt reorganization improves the convergence of the successive approximations at large coupling, $e \sim 2$. We also compared the HTLpt three-loop result with a three-loop $\Phi$-derivable approach\cite{phijm} and found agreement at the subpercentage level.

In closing, we mention that the corresponding three-loop calculation for pure-gauge QCD was just completed\cite{Andersen:2009tc} and the resulting thermodynamic quantities are consistent with lattice data\cite{Boyd:1996bx} down to temperatures $T\sim2-3~T_c$ which indicates that the lattice data at these temperatures are consistent with the picture of a plasma of weakly-coupled quasiparticles. Since HTLpt is formulated in Minkowski space, its application to real-time dynamics could be important to heavy-ion phenomenology.

\begin{center}
\vspace{10mm}
\includegraphics[width=7.5cm]{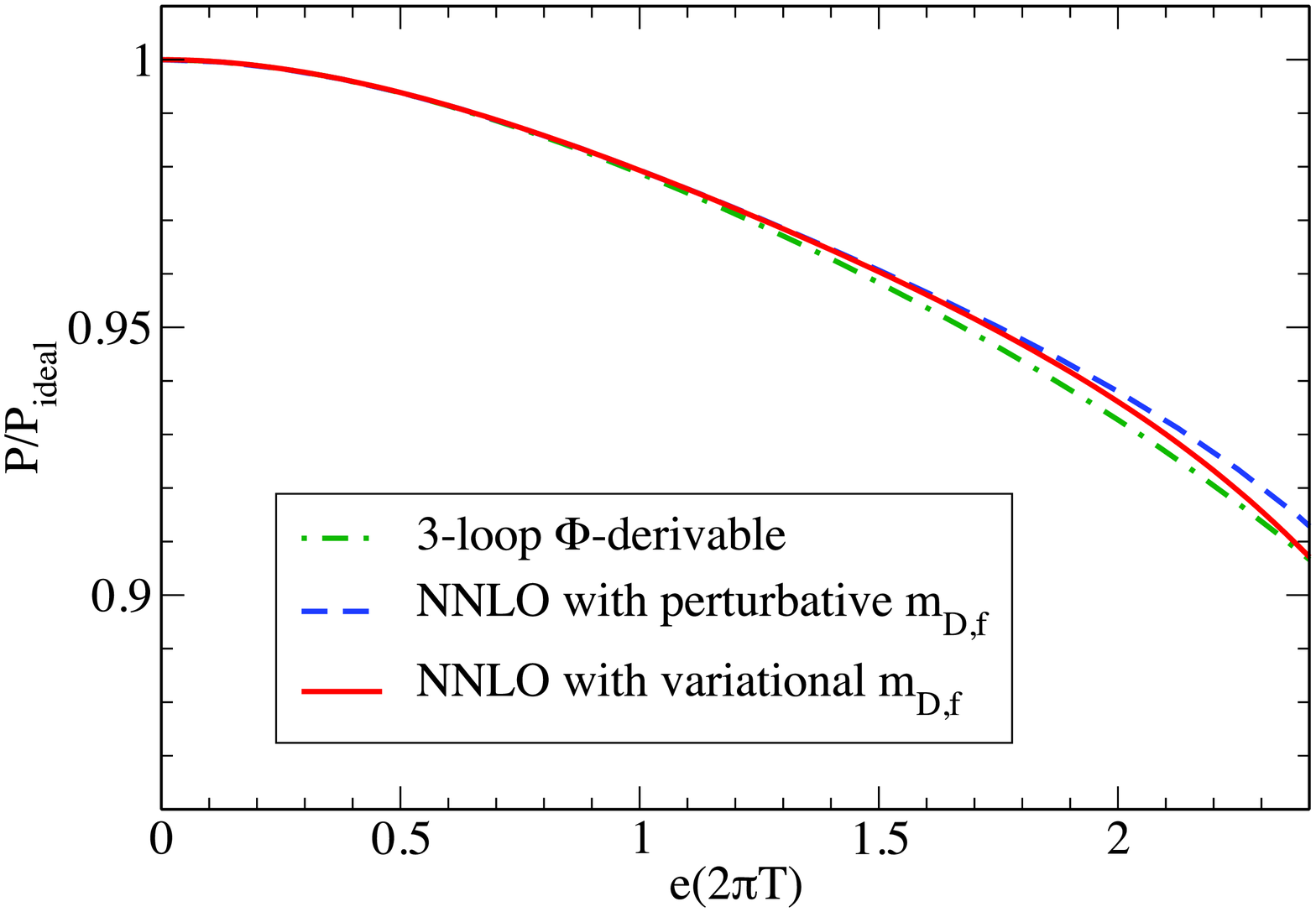}
\figcaption{\label{fig:PhivsNNLO} A comparison of the predictions for the free energy of QED with $N_f=1$ between three-loop $\Phi$-derivable approximation and NNLO HTLpt at $\mu=2\pi T$.}
\end{center}

\acknowledgments{N. S. was supported by the Frankfurt International Graduate School for Science. M. S. was supported in part by the Helmholtz International Center for FAIR Landesoffensive zur Entwicklung Wissenschaftlich-\"Okonomischer Exzellenz program.}

\end{multicols}

\vspace{-2mm}
\centerline{\rule{80mm}{0.1pt}}
\vspace{2mm}

\begin{multicols}{2}

\end{multicols}

\clearpage

\end{document}